\documentclass[structabstract]{aa}  
\usepackage{natbib}
\usepackage{graphicx}
\usepackage{amssymb}
\usepackage{txfonts}
\begin{document}
   \title{Time-dependent escape of cosmic rays from supernova remnants, and 
their interaction with dense media}

   \subtitle{}

   \author{I.Telezhinsky
          \inst{1,2}\fnmsep\thanks{\email{igor.telezhinsky@desy.de}}
          \and
	  V.V. Dwarkadas
          \inst{3}
	  \and
          M. Pohl
          \inst{1,2}
          }

   \institute{DESY, Platanenallee 6, 15738 Zeuthen, Germany
         \and
            Universit\"{a}t Potsdam, Institut f\"{u}r Physik \& Astronomie, Karl-Liebknecht-Strasse 24/25, 14476 Potsdam, Germany 
         \and            
             University of Chicago, Department of Astronomy\&Astrophysics, 5640 S Ellis Ave, AAC 010c, Chicago, IL 60637, U.S.A. 
             }

   \date{Received ; accepted }


  \abstract
{Supernova remnants (SNRs) are thought to be the main source of
  Galactic cosmic rays (CRs) up to the ''knee'' in CR spectrum. During the
    evolution of a SNR, the bulk of the CRs are confined inside the
    SNR shell. The highest-energy particles leave the system
    continuously, while the remaining adiabatically cooled particles
    are released when the SNR has expanded sufficiently and decelerated so that the magnetic field at the shock is no longer
    able to confine them. Particles escaping from the parent system
    may interact with nearby molecular clouds,
    producing $\gamma$-rays in the process via pion decay. The soft
    gamma-ray spectra observed for a number of SNRs interacting with
    molecular clouds, however, challenge current theories of
    non-linear particle acceleration that predict harder spectra.}
{We study how the spectrum of escaped particles depends on the
  time-dependent acceleration history in both Type Ia and
  core-collapse SNRs, as well as on different assumptions about the
  diffusion coefficient in the vicinity of the SNR.}  {We solve the CR
  transport equation in a test-particle approach combined with
  numerical simulations of SNR evolution.} {We extend our method for
  calculating the cosmic-ray acceleration in SNRs to trace the escaped
  particles in a large volume around SNRs. We calculate the evolution
  of the spectra of CRs that have escaped from a SNR into a molecular cloud or
  dense shell for two diffusion models. We find a strong confinement
  of CRs in a close region around the SNR, and a strong dilution
  effect for CRs that were able to propagate out as far as a few SNR
  radii.}
   {}

\authorrunning{I. Telezhinsky et al.}
\titlerunning{Time-dependent escape of CRs from SNRs, and their Interaction with dense media}

\maketitle
%

\section{Introduction}

Supernova remnants (SNRs) are now widely considered to be sources of Galactic
cosmic rays (CRs). Diffusive shock acceleration (DSA) \citep{Axfetal77,
  Kry77, Bel78, BlaOst78} and its non-linear modification (NDSA)
\citep{MalDru01} predict a power-law distribution ($N(E)\propto
E^{-s}$) of relativistic particles with index $s=2$ ($s<2$ for the
high-energy tail in NDSA). Despite recent advances in DSA and NDSA, it
is still unclear why we observe soft ($s>2$) CR spectra at Earth
\citep{Aveetal09} and soft gamma-ray spectra from a number of SNRs:
RX~J0852.0-4622 \citep{Ahaetal07b}, RCW~86 \citep{Ahaetal09}, SN~1006
\citep{Aceetal2010_1006}, Cas~A \citep{2010ApJ...714..163A,
  Abdetal10_CasA}, and Tycho's SNR
\citep{Accetal11,Gioetal11}.

Understanding gamma-ray SNR spectra in various situations
  requires the study of the acceleration of the particles in the system
  (consisting of one or two shock waves), followed by the escape of
  these particles, their diffusion into the Galactic medium
  \citep{Ptuetal10, Capetal10}, and computation of the $\gamma$-ray
  emission by various processes.  In an earlier paper
  \citep{Teletal11}, we studied the acceleration of particles at
  SNR shock fronts. In this paper, we follow up this work by studying
  the escape of these particles from the system, and their interaction
  with dense media.

A useful probe of the escape of particles from
  SNRs may be high-energy gamma-ray emission from molecular clouds
located close to an efficiently accelerating SNR, the so-called MC-SNR
systems. A number of these have been observed by Cherenkov
telescopes (CTs) and \textit{Fermi}. Most sources, including W28
\citep{Abdetal10_w28}, W44 \citep{Abdetal10_w44}, IC~443
\citep{2009ApJ...698L.133A, Abdetal10_ic443}, G349.7+0.2, CTB 37A,
3C391, and G8.7-0.1 \citep{CasSla10} are located close to dense
regions or molecular clouds (MCs), thus one may use them to infer the diffusion
parameters in their vicinity. Triggered by these observations, a
number of studies \citep{Gabetal09, Fujetal09} have attempted to
explain the soft gamma-ray emission from MC-SNR systems in terms of escaped
CRs. These studies have been based on analytical models \citep{Atoetal95, AhaAto96} and several assumptions: (i) the SNR evolution is either stationary or Sedov-like, (ii) often that particle acceleration is quasi-instantaneous compared with the CR diffusion time, and (iii) the escaping-particle distribution is a power-law
or monoenergetic at a given maximum momentum.  The source of CRs is
considered to be point-like, i.e., the SNR radius is much smaller than
the distance to the MC. \citet{LiChe10}, \citet{LiChe11}, and \citet{Ohietal11},
extended the analytical models to include finite-size sources and
finite-size target MCs.

Despite their simplicity, the analytical studies demonstrate that to
explain observations, the diffusion coefficient in the SNR vicinity
must be roughly two orders of magnitude smaller than the average
Galactic value, supporting earlier claims \citep{Wen74} that the
diffusion coefficient might be smaller because of plasma
waves that scatter particles. Using a simple model of SNR evolution
and Monte Carlo simulations of CR diffusion, \citet{Fujetal10} showed
that particles may be trapped around the SNR for a significant period of
time. Subsequently, \citet{Fujetal11} confirmed this result using a
transport equation to describe particle acceleration at the forward
shock (FS), and then a simplified transport equation to follow the
propagation of particles with $p/m_0c \geq 10^{2.4}$ from the
precursor boundary (set at four Bohm diffusion length-scales) to
distances far away from the shock. These findings are consistent with
the conclusions of \citet{Revetal09}, who argue that beyond some
critical value, $L$, which they call the free-escape boundary, the
number density of CRs falls significantly. The position of $L$ is
dictated by the level of excited magnetohydrodynamics (MHD) turbulence, which may be
significant out to one SNR radius from the FS \citep{ZirPtu08a}.  The
detailed variation of the turbulence amplitudes and the CR diffusion
coefficient depends on the micro-physical balance of streaming
instabilities and turbulence damping \citep{Yanetal12}.

Besides the diffusion coefficient in the SNR vicinity, other important
aspects of the study of escaped CRs are their spectral and spatial
distributions. These distributions affect the radiative properties of
the SNR itself as escaping CRs modify the high-energy part of the CR
spectrum, and, if present, the radiative properties of the nearby
dense matter consisting of a MC or a dense shell swept up by the winds of a
high-mass SN progenitor). Much of the research dealing with stationary
and (semi-)analytical solutions cannot recover the spectral and
spatial shapes of the CR distribution, but only gives the integrated
escaping CR energy flux. Kinetic models provide the shape of the
distribution, which depends significantly on the assumed diffusion
models.  \citet{EllByk11} parametrized the spectral distribution of
escaping CRs, and then propagated the escaping CRs in the upstream
region of the SNR obtaining spatial distributions. They used a
spherically symmetric code for the hydro simulations of the SNR
evolution that is identical to the code used in this work, but
without the grid expansion that significantly increases the
resolution. However, the particle acceleration was treated in a
plane-parallel steady-state approximation ignoring dilution
effects. \citet{EllByk11} focused on the core-collapse
explosion of a massive star in a low density medium surrounded by a
dense shell, a case similar to our model for a core-collapse SNR (CC-SNR) in a
wind bubble.

In this paper, we investigate how both the spectrum of particles
  escaped from the SNR into a nearby MC or shell and the particle emission
due to interaction with dense matter depends on the
acceleration history of young ejecta-dominated Type-Ia and Type-Ic (core-collapse) SNRs, using given
different assumptions about CR diffusion in the vicinity of the
SNR. For this purpose, we combine a test-particle treatment of CR
acceleration with an account of escaping particles. We solve the CR
transport equation in a spherically symmetric geometry. Our calculations
are based on realistic high-resolution hydrodynamic simulations of SNR
evolution. We explore two SN types, focusing on the complexities in
the circumstellar environment. In the case of Type-Ia SNRs, we assume that 
evolution proceeds in a uniform ISM. We also consider a core-collapse SNR
arising from a Wolf-Rayet (WR) progenitor, expanding in the wind-blown
bubble created by the progenitor star.  The SNR shock wave first
evolves in the freely expanding wind of the progenitor, and then,
beyond the wind-termination shock, in the shocked-wind
region. Subsequently, we would expect the shock to impact the dense
shell bordering the wind-blown bubble. However, our intention here is
to study the interaction of the accelerated particles produced by the
SNR with this dense shell, hence we terminated the SNR
evolution before the shock-shell impact took place. We considered
particle acceleration at both the forward and reverse
shock. In addition, in the CC-SNR case we consider
particle re-acceleration at secondary shocks formed when the SNR
forward shock collides with the termination shock of the wind
zone. The spatial domain considered in our particle
  simulations extends to about 100 times the SNR radius to facilitate
  the study of particle escape. The ``absorbing'' boundary located
  that far from the SNR does not affect the result of our
  simulations. It is the spatial profile of the diffusion
  coefficient in the vicinity of the SNR that determines the degree to which
 particles can escape from the CR precursor.  To estimate the
observational consequences of CR escape, we consider the interaction
  of the escaped particles with a target MC close to a Type-Ia SNR, 
or in the core-collapse case we consider the interaction
of CRs with the dense shell. The MC/shell are located well
  inside the simulation domain, hence we know the cosmic-ray
  number density at their location at any time. Using two different
models of the CR diffusion upstream of the FS, we calculated the evolutions
of both the cosmic-ray spectra at the location of the MC/shell and the
hadronic gamma-ray emission as the SNR shock approaches the MC/shell. 
If not stated otherwise, we refer to particles as escaped
  when they have escaped from the precursor region of the FS
  and are found at the location of the MC or shell. The respective
  particle spectra are averaged over the volume of the MC or
  shell.

\section{Method}

We describe the hydrodynamics of the expanding SNR by numerical
simulations. For the Type-Ia SNR, we use the ejecta-density profile
described in \citet{DwaChe98}, expanding into a constant-density
medium. The simulation itself is described in detail in
\citet{Teletal11}. The evolution of core-collapse Type-Ic SNRs is much
more complicated. We assume that the formation of a wind-blown bubble is similar to that
described by \citet{Dwa05, Dwa07}. The WR wind is
assumed to have a mass-loss rate of $ 10^{-5} M_{\odot} {\rm
  yr}^{-1}$, and a velocity of 1500 km s$^{-1}$. The wind termination
shock is assumed to be a strong shock that forms at a radius of 7
pc. As the SNR shock expands outwards in radius, it evolves in
the freely expanding wind, impacts the wind termination shock, and
then evolves into a constant density shocked wind. Finally, although
this is not included in the current simulation, the shock will collide with
the dense wind-blown shell surrounding the bubble.

We treat CRs as test particles in gas-flow profiles given by the
simulations described above. Our method \citep{Teletal11} is based on a
numerical solution of the CR transport equation in a grid co-moving
with the shock wave. To ensure sufficient resolution near the shock,
the spatial coordinate is substituted with a new coordinate, $x_*$,
for which a uniform grid is used when solving the particle transport
equation
\begin{equation}
(x-1)=\left(\frac{r}{R_{sh}} -1\right)=(x_*-1)^3.
\label{transform}
\end{equation}
Thus, a rather coarse grid in $x_*$ is transformed into a very fine
grid in $x$ close to the shock, where the high resolution is needed to
properly account for the acceleration of newly injected particles. At
the same time, this transformation allows us to significantly extend
the grid far into the ISM ($x\gg 1$) with only a small extension in
$x_*$. The resolution obviously deteriorates with distance from the
shock, but the mean free path of the particles in the CSM/ISM is
orders of magnitude larger than in the shock vicinity, thus permitting
the use of a moderate resolution. The extension of the grid to several
dozens of SNR radii, and the spherical geometry, imply that
  the volume upstream of the shock is much larger than downstream. The co-moving grid obliterates any need to introduce an
  artificial ``absorbing'' or ``escape'' boundary because the particles
  will never reach it during the simulation time. The number of particles escaping
from the SNR shock is defined purely by the diffusion properties of
the upstream region. We do not need to define the diffusive
  flux through some artificial boundary to count the
  particles escaping through this boundary. In the current approach,
  owing to the grid extension the total number of injected particles is
  conserved and distributed over
  the considered volume according to diffusion. By the end of the
  simulation, we observe at the boundary of the simulation domain only
  numerical noise, which indicates that the boundary is reached by an insignificant fraction of
  particles. Technically, a macroscopic treatment of
  diffusion does not differentiate between individual particles
  bouncing back and forth in the shock precursor, thus all particles
  found upstream of the FS can be assumed to have escaped from the
  SNR\footnote{This is not true in a microscopic treatment, where any
    individual particle has a non-zero probability of returning back
    to the shock from any distance.}. Here we are interested in
  particles found at the location of the MC/shell (sufficiently far away from
  the shock precursor), and so we refer to these particles as escaped.

Since our method is based on the test-particle approximation, it
requires that the CR pressure at the shock be less than 10\% of the
ram pressure. We use a thermal-leakage injection model
\citep{Blaetal05}, and adjust the injection so that the CR-pressure limit
is not violated. The injection coefficient, a free parameter, is
adjusted to be approximately $3\cdot10^{-7}$ for the Type-Ia SNR, and
$5\cdot10^{-6}$ for CC-SNR.

\section{Magnetic field and diffusion}
\subsection{Magnetic field parametrization}

Observations of young SNe, for example SN 1993J \citep{Chaetal04} or
Tycho's SNR \citep{Accetal11}, suggest that the magnetic field (MF)
required to explain the observed level of emission is orders of
magnitude stronger than the average ISM field. We therefore assume
that the MF in the shock vicinity is amplified, e.g. by streaming CRs
\citep{Bel04}. Although in general both resonant and non-resonant
modes operate, non-resonant amplification probably dominates in the
early stages of SNR evolution \citep{Capetal09b}. The amplified field
is then given by
\begin{equation}
B_0(t) = \sqrt{2 \pi \rho_u(t) \left(\frac{V_{s}(t)^3}{c}\right) \xi(t)},
\end{equation} 
where $\rho_u(t)$ is the density upstream the shock, $V_s(t)$ is the
shock speed, $c$ is the speed of light, and $\xi(t)$ is the ratio of
cosmic-ray to ram pressure. In our calculations, $\xi(t) \approx 0.05$
with small deviations throughout the simulation.

We parametrize the MF inside and outside the SNR. The scaling inside
follows the time-dependent density distribution
\begin{equation}
B(r,t) = \sigma \frac{B_0(t) \rho(r,t)} {\rho(R_{FS},t)},
\end{equation}
where $\sigma=\sqrt{11}$ is the compression ratio of the turbulent MF,
and is assumed to be isotropic. We assume that the MF falls off
exponentially down to the strength of the interstellar field (5~$\mu$G) (or circumstellar
MF (CMF) in the case of core-collapse SNR), at $0.05\,R_{\rm FS}$ ahead of
the FS \citep{ZirAha10}, and likewise down to the very small ejecta
field (0.01-0.1~$\mu$G) at $0.05\,R_{\rm RS}$ toward the interior of
the SNR.

\begin{figure}[!t]
\vspace{5mm}
\centering
\includegraphics[width=0.48\textwidth]{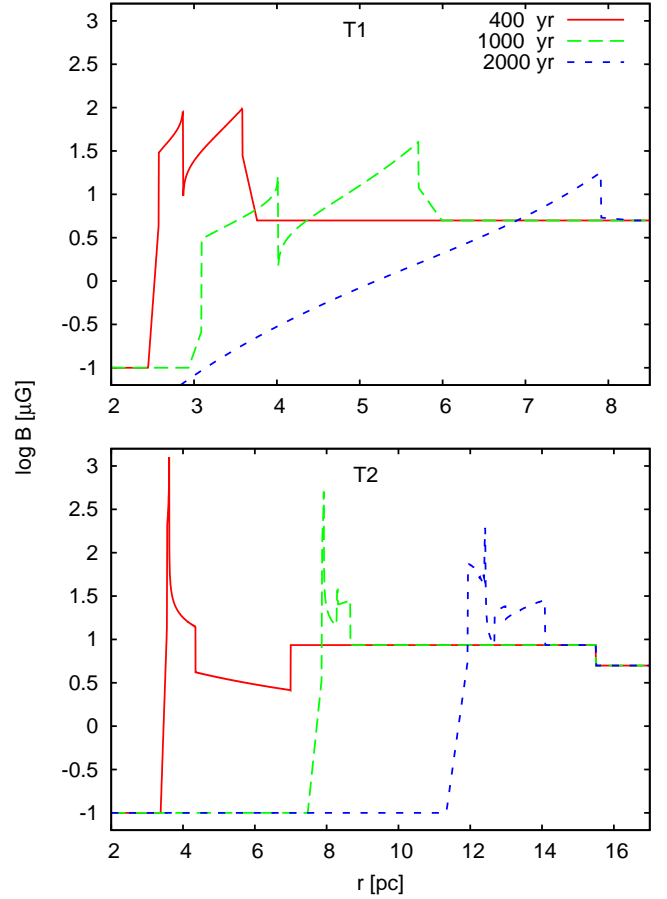}
\caption{Time dependence of magnetic-field profiles for type-Ia (top)
  and core-collapse (bottom) SNRs.}
\label{mf}
\end{figure}

To define the magnetic field in the wind zone of the WR star, we
assume that the dominant component at large distances is merely the
toroidal component of the stellar surface magnetic field, which
decreases outwards with radius. Thus, this component is defined as
\begin{equation}
B_{c}(R) = \frac{B_s R_{\rm WR}}{R},
\end{equation}
where $B_s\approx 100$~G is the MF at the surface of WR star, $R_{\rm
  WR}=8R_{\odot}$ is the radius of WR star, and $R$ is the distance
from the star. Beyond the WR wind zone, in the constant density
region, the MF is assumed to be constant and equal to the
shock-compressed value at the edge of the WR wind zone.
We assume a constant MF in the MC/shell, since it has been
found that the MF does not increase if the number density in the cloud, $n \lesssim
300$cm$^{-3}$ \citep{Cruetal10}. The evolution of the MF profiles for
two types of SNR is shown in Fig.~\ref{mf}.

\subsection{Diffusion models}

\begin{figure}[!b]
\vspace{5mm}
\centering
\includegraphics[width=0.48\textwidth]{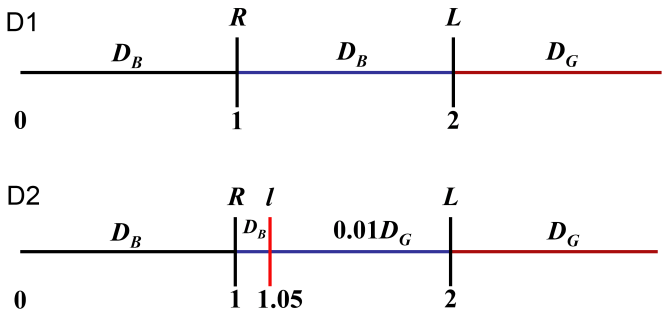}
\caption{Schematic view of diffusion models.}
\label{diffm}
\end{figure}

To study the propagation of cosmic rays in the vicinity of
SNRs, it is necessary to make assumptions about the efficiency of diffusion
close to the SNR shock. We note that the
type of particle scattering does not depend on the amplitude of the
MF, but rather on the wave-number spectrum of MHD turbulence. In other
words, one can have Bohm-type diffusion in the non-amplified MF even
significantly far away from the shock. Bohm diffusion is widely assumed
to operate close to the shock and inside the SNR, otherwise DSA
theories would fail to explain particles with energies up to the
''knee''. On the other hand, the average diffusion coefficient in the
Galaxy is much larger than Bohmian, and inevitably there must be a
transition between the small Bohmian coefficient and the large
Galactic coefficient.

Since the transition from Bohm diffusion in the vicinity of the
forward shock to Galactic diffusion further away is not clearly understood, we
consider two different models of CR diffusion outside the SNR. In
model D1, the diffusion coefficient is Bohmian inside the SNR and in
the FS upstream region, up to a fiducial boundary, $L=2R_{SNR}$, at
which the transition to the Galactic diffusion occurs. We emphasize
  that the boundary L is not a physical boundary in any sense, and
  does not denote an ``escape'' boundary as in our previous work. It is merely a convenient radius at which the transition in
  diffusion coefficient is assumed to occur. In this model the
diffusion coefficient is defined as
\begin{equation}
D(r) = \left\{ \begin{array}{ll}
      D_B	& \textrm{ $r \leq L$},\\
      D_G	& \textrm{ $r > L$},
\end{array} \right.
\label{eqd1}
\end{equation}
where the Bohm diffusion coefficient is given by
\begin{equation}
D_B = \frac{pvc} {3qB}
\end{equation}
while the Galactic diffusion coefficient is taken to be
\citep{Beretal90}
\begin{equation}
D_G=D_0 \left(\frac{E} {10\textrm{ GeV}}\right)^{\alpha} \left(\frac
{B} {3\mu\textrm{G}}\right)^{-\alpha} \textrm{cm$^2$/s,}
\end{equation}
where $E$ is the CR energy and $D_0$ is the normalization. Following a 
statistical analysis of cosmic-ray propagation models
\citep{Troetal11}, we use $D_0 = 10^{29}$cm$^2$/s and $\alpha = 1/3$.

In model D2, we explore a less abrupt transition to Galactic
diffusion. We assume that Bohm diffusion is valid both inside
the SNR and in the FS upstream region stretching out to 5\% of the SNR
radius, i.e., up to $l=1.05 R_{SNR}$. Between $l$ and $L$, we introduce
an intermediate diffusion coefficient, smaller than the
Galactic diffusion coefficient, thus mimicking the presence of MHD waves,
which may be invoked by CRs streaming away from the SNR \citep{Yanetal12}.
We define the diffusion coefficient in D2 model as
\begin{equation}
D(r) = \left\{ \begin{array}{ll}
      D_B	& \textrm{ $R_{SNR} \leq r \leq l$},\\
      \chi D_G	& \textrm{$l < r \leq L$},\\
      D_G	& \textrm{ $r > L$},
\end{array} \right.
\label{eqd2}
\end{equation}
where $\chi=0.01$. A schematic view of the diffusion models and 
corresponding boundaries is given in Fig.~\ref{diffm}.

We do not consider the confinement of particles by the dense shell at
the boundary of the WR wind zone in the CC-SNR models as done by
\citet{EllByk11}. The ion-neutral friction in dense and partially
ionized media may suppress plasma instabilities, and thus increase
the diffusion coefficient \citep{PtuZir05,Dru11}. Similar
  arguments would apply to MCs. For instance, \citet{EveZwe11}
  consider the diffusion coefficient in MCs to be 1\%--500\%
  of the Galactic value. They claim that this variation does not
  significantly affect their findings, which state
that the cosmic-ray density does not increase within
clouds. Therefore, we assume that the diffusion coefficient
inside the MC/shell is the same as in the surrounding
medium. A possible variation of the diffusion coefficient inside the
  MC/shell and the resulting consequences for our calculations are
  discussed in subsection \ref{pspe}.

\section{Results and discussion}

\begin{figure*}[!t]
\centerline{\includegraphics[width=0.97\textwidth]{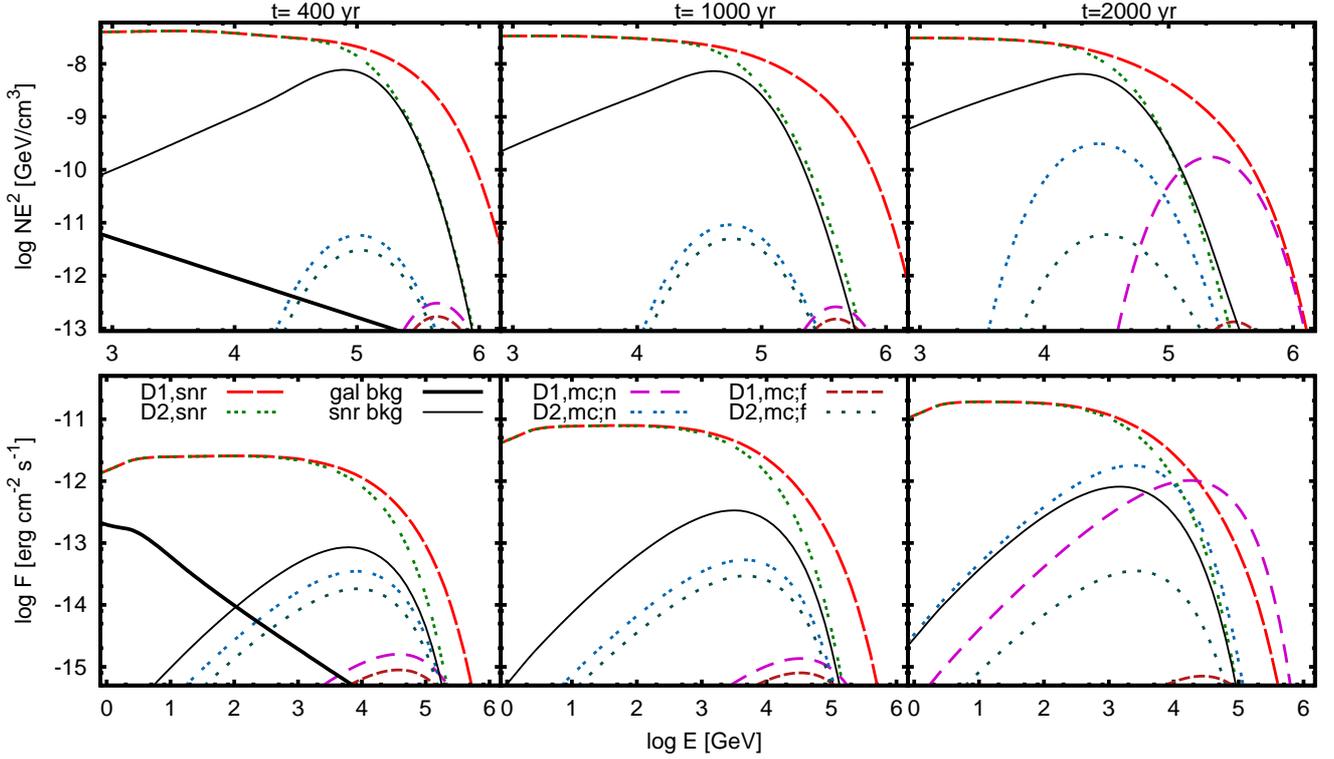}}
\caption{Results for Type-Ia SNR. Top: the time evolution of
  volume-integrated CR spectra inside the SNR (denoted ``snr'') and
  the MC (denoted ``mc'') for two different diffusion models
  (``D1''/``D2'') and two different distances from the SNR to MC:
  12~pc (``n'') and 16~pc (``f''). Bottom: the corresponding hadronic
  gamma-ray emission spectra.  The thick solid lines represent the
  Galactic CR background and the corresponding gamma-ray emission from
  the MC, both of which do not depend on time.  The thin solid lines
  represent the spectra of CRs outside the SNR for the D2 model and
  the corresponding emission from ambient diffuse gas.}
\label{T1}
\end{figure*}

We have calculated the confined and escaped cosmic-ray spectra at 400,
1000, and 2000 years after explosion, assuming the diffusion models
described above. We have also calculated the corresponding emission
from the SNRs and the target material in their vicinity. In
calculating the emission, we considered only CR protons, and their
radiation via pion decay \citep{Huaetal07}.

Given the initial parameters assumed in the simulation, the Type-Ia
SNR approaches the Sedov stage after around 1000 years. Beyond this
time, we approximate the plasma-flow profiles with a Sedov solution
\citep{CoxFra81}. Although hydrodynamically the RS is still present,
usage of the Sedov solution is justified since the contribution from
the RS to particle and emission spectra is negligible at that time
\citep{Teletal11}. After 2000 years, the FS radius in Type-Ia SNR is
about 8~pc. The CC-SNR, on the other hand, by design, expands
in a wind of much lower density, followed by a constant-density
shocked-wind zone, and therefore remains in the ejecta-dominated stage
until the end of the simulation. At the edge of the constant-density
zone, we assume the presence of a cold, thin (0.5~pc) dense shell of
swept-up material, similar to those known to exist around WR stars
\citep{capetal03}. After 2000 years, the forward shock of this Type-Ic
SNR has expanded out to a radius of about 14~pc.

The spectrum of particles interacting with the MC is a function of its
distance from the SNR shock, as well as the diffusion coefficient. We
consider two center-to-center distances from the Type-Ia SNR to the
MC, 12~pc (the ``near'' scenario) and 16~pc (the ``far'' scenario),
thus study the effect of a variation in the diffusion coefficient on
  the particle spectrum as the cloud approaches the SNR. The MC is
{assumed to have a radius} of 4~pc and a mass of $M_{c}\approx
1500\,$M$_{\odot}$, corresponding to a number density $n_{c} =
150\,$~cm$^{-3}$.

For the CC-SNR case, we consider a swept-up shell located at 16~pc (the
``near'' scenario) or at 30~pc (the ``far'' scenario). The number
density of the shell, $n_{s} = 100\,$~cm$^{-3}$, corresponds to a mass
$M_{s} \approx 1300\,$M$_{\odot}$ for the ``near'' and $M_{s} \approx
4500\,$M$_{\odot}$ for the ``far'' scenario. We note that the gamma-ray
flux simply scales with the gas mass in the MC or the dense shell, and
therefore the choice of mass only affects the normalization of the
spectra.

\subsection{Particle spectra}
\label{pspe}

\begin{figure*}[!t]
\centerline{\includegraphics[width=0.97\textwidth]{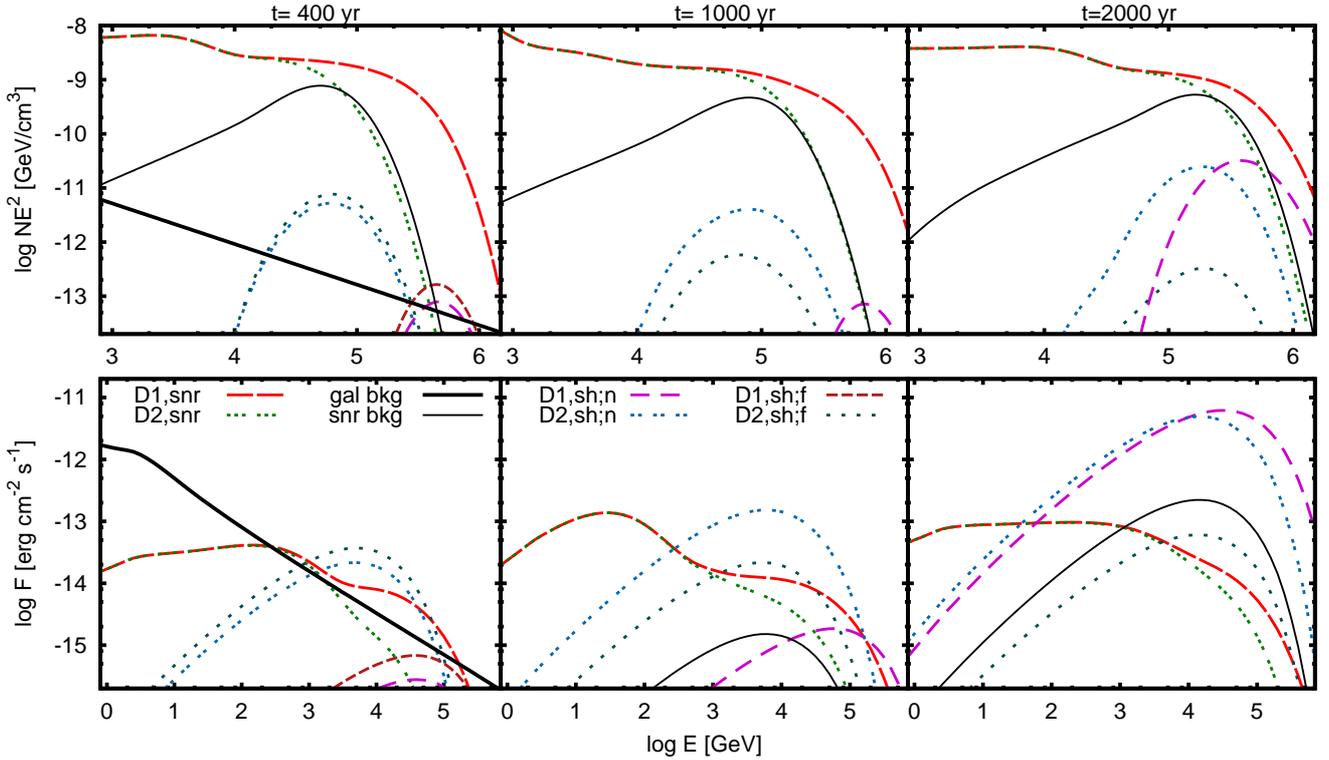}}
\caption{Results for core-collapse SNR. Top: the time evolution of
  volume-integrated CR spectra inside the SNR (denoted ``snr'') and at
  the dense shell (denoted ``sh'') for two different diffusion models
  (``D1''/``D2'') and two different distances from the SNR to the
  shell: 16~pc (``n'') and 30~pc (``f''). Bottom: the corresponding
  hadronic gamma-ray emission spectra.  The thick solid lines
  represent the Galactic CR background and the corresponding gamma-ray
  emission from the dense shell, both of which still do not depend on
  time.  The thin solid lines represent the spectra of CRs outside 
  the SNR for the D2 model and the corresponding emission from ambient
  diffuse gas.}
\label{T2}
\end{figure*}

The time evolution of the spectra of confined and escaped protons are
presented in the upper panels of Fig.~\ref{T1} for Type-Ia SNR and
Fig.~\ref{T2} for CC-SNR, respectively. All spectra are
volume-averaged. The Galactic CR background and its gamma-ray
emission are shown for comparison.

We note from the figures that the choice of diffusion model
influences the cut-off energy in the spectra of the confined
particles, as well as the energy of the spectral peak of escaped CRs.
Particle acceleration becomes slow when the CR precursor extends to
the radius $l$ in the upstream region at which diffusion becomes
faster in the D2 model. Thus, the spectra of confined particles cut
off at lower energies, and likewise the spectra of escaped CRs have
maxima at lower energies, because escape from the SNR is more
efficient.  In both models of diffusion and both types of SNR, we do
not observe any significant contribution of the RS to the population
of the CRs found upstream of the FS. This is partly because
the reverse-shock contribution diminishes with time, and is largely
negligible after about 1000 years. Furthermore, since the MF is
assumed to scale with the SNR density in the shocked interaction
region, the large MF at the contact discontinuity, and the low maximum energy attained by CRs at the
reverse shock do not permit particles accelerated at the RS to diffuse
out of the SNR.

In Type-Ia SNR, the spectral shapes of escaped particles in the D1 and D2
diffusion models are similar. They have a narrow, parabola-like shape
at a certain energy, usually referred to as $E_{\rm max}$, that is consistent with 
the approximations made by \citet{EllByk11}. The
intensity changes with time depending on the assumed distance to the
MC. In the ``far'' scenario, the MC always stays beyond the $L$
boundary, and the intensity of the CR spectra at the location of the
MC rapidly decreases with time on account of dilution. In the
``near'' scenario, the MC is partially (after 1000 years) or totally
(after 2000 years) within $L$, where the number density of CRs is
high. However, the CRs in the D2 model have a shallower gradient than in the
D1 model, and so the CR illumination of the MC increases even if the
MC is only partially within $L$ (1000 years). In the D1 model, the
spectrum is still affected by dilution, and it is only later (after
2000 years), when the MC is totally inside $L$, that the spectral
intensity rises dramatically.

\begin{figure*}[!t]
\centerline{\includegraphics[width=0.95\textwidth]{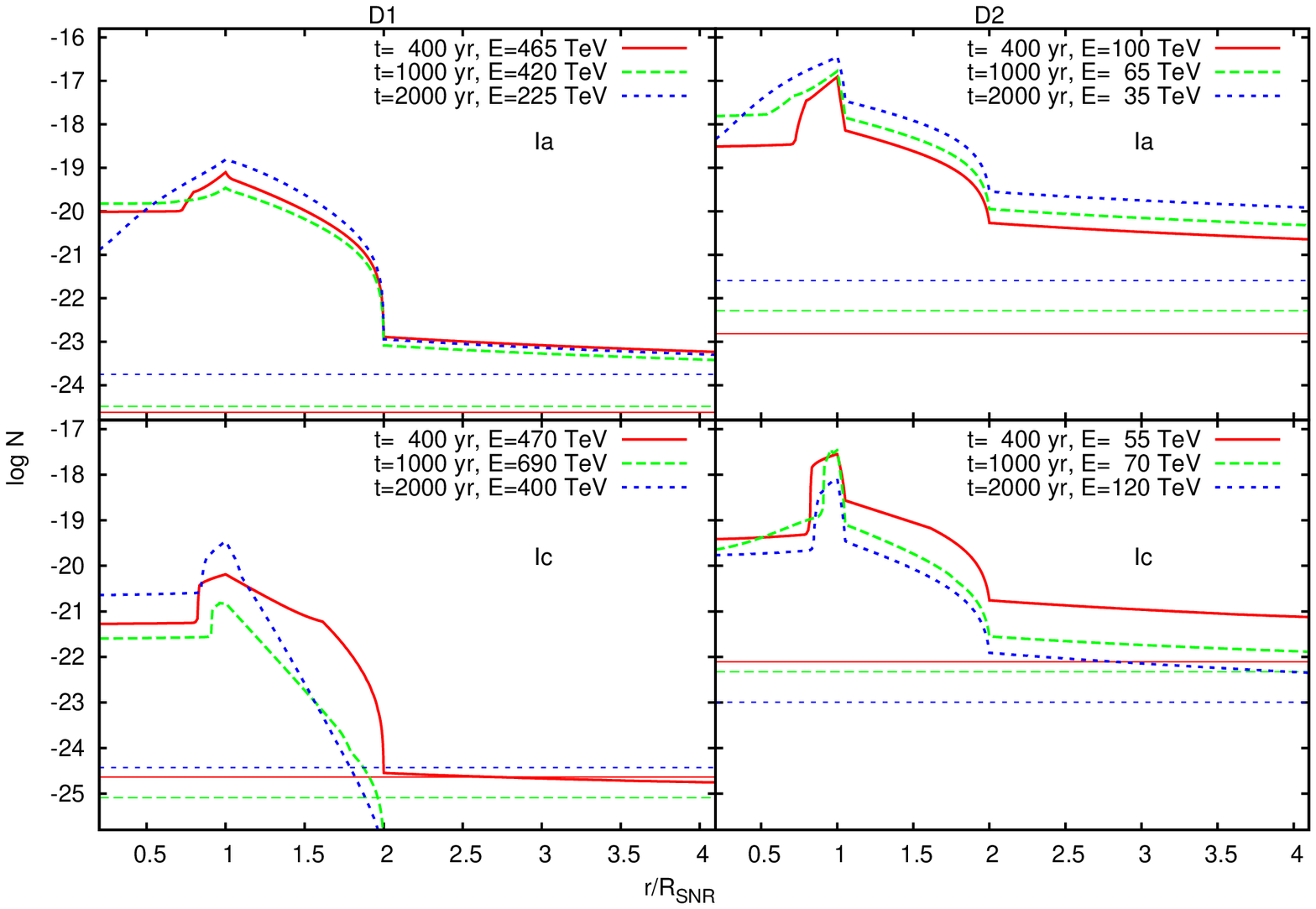}}
\caption{Radial distributions of CRs at the peak energy
in the spectra of escaped CRs at the given time
(thick lines), compared with the Galactic CR background 
at the same energies (thin lines) for diffusion models ``D1'' (left)
and ``D2'' (right). Type-Ia SNRs are shown in the top row and
core-collapse SNR in the bottom row.}
\label{nar}
\end{figure*}

The behavior of escaped particles in the D1 model is similar for both
types of SNR. However, the spectral shape of the escaped particles in
the D2 model is somewhat different in the CC-SNR case. The spectral
peak for the escaped CRs shifts to higher energies with time. Therefore, an asymmetry is seen
in their spectra, most visibly so at later times.

We now discuss the consequences of a different diffusion
  coefficient inside the MC/shell. First, we note that the shell
  thickness is so small that the shell spectra are virtually
  unaffected, unless the diffusion coefficient inside is extremely
  small. Suppose then that the plasma waves are damped so
  effectively that the diffusion coefficient inside the MC increases
  significantly (about ten times) with respect to the diffusion
  coefficient of the surrounding medium. It is obvious that the CRs
  will quickly attain a uniform distribution inside the MC, and the CR
  number density at the far side of the MC will be nearly equal to
  that at the near side. This should affect the CR intensity inside
  the MC after about 1000~years, when the MC is located at a position corresponding to a strong
  gradient in the CR distribution (it is partially within boundary
  $L$). Estimates show that the CR intensity in this case may
  increase by a factor of a few. At other times, the differences will
  be marginal.

The diffusion coefficient inside the MC may also be lower than
  in the ISM.  This may be due to a strong gradient in CRs at the
  outskirts of the MC, which may trigger plasma instabilities and in
  turn reduce the CR diffusivity. A reasonable value could be $0.01
  D_{G}$. We note that this case may affect the D1 model
  after 1000 years in the same manner as described above, because the
  diffusion coefficient inside the MC will be larger than within the
  boundary $L$. Spectra at other times and for other models should not be
  affected.  The main effect of a reduced diffusion coefficient inside
  the MC is the longer penetration time and consequently a delay in
  the illumination of the MC. If one estimates the time needed for a
  10-TeV particle to reach the center of the MC, one finds
  $t=R_{MC}^2/2D_{MC} \simeq 230$~years, where $D_{MC} =
  0.01D_{G}$. However, the outer 1.5-pc thick shell of the MC would then supply
  80\% of the flux, and a 10-TeV CR would need only $\sim$30 years to
  penetrate to this depth. The delay in illumination is therefore
  negligible compared with the uncertainties in the size and structure of
  the MC.

\begin{figure*}[!t]
\centerline{\includegraphics[width=0.95\textwidth]{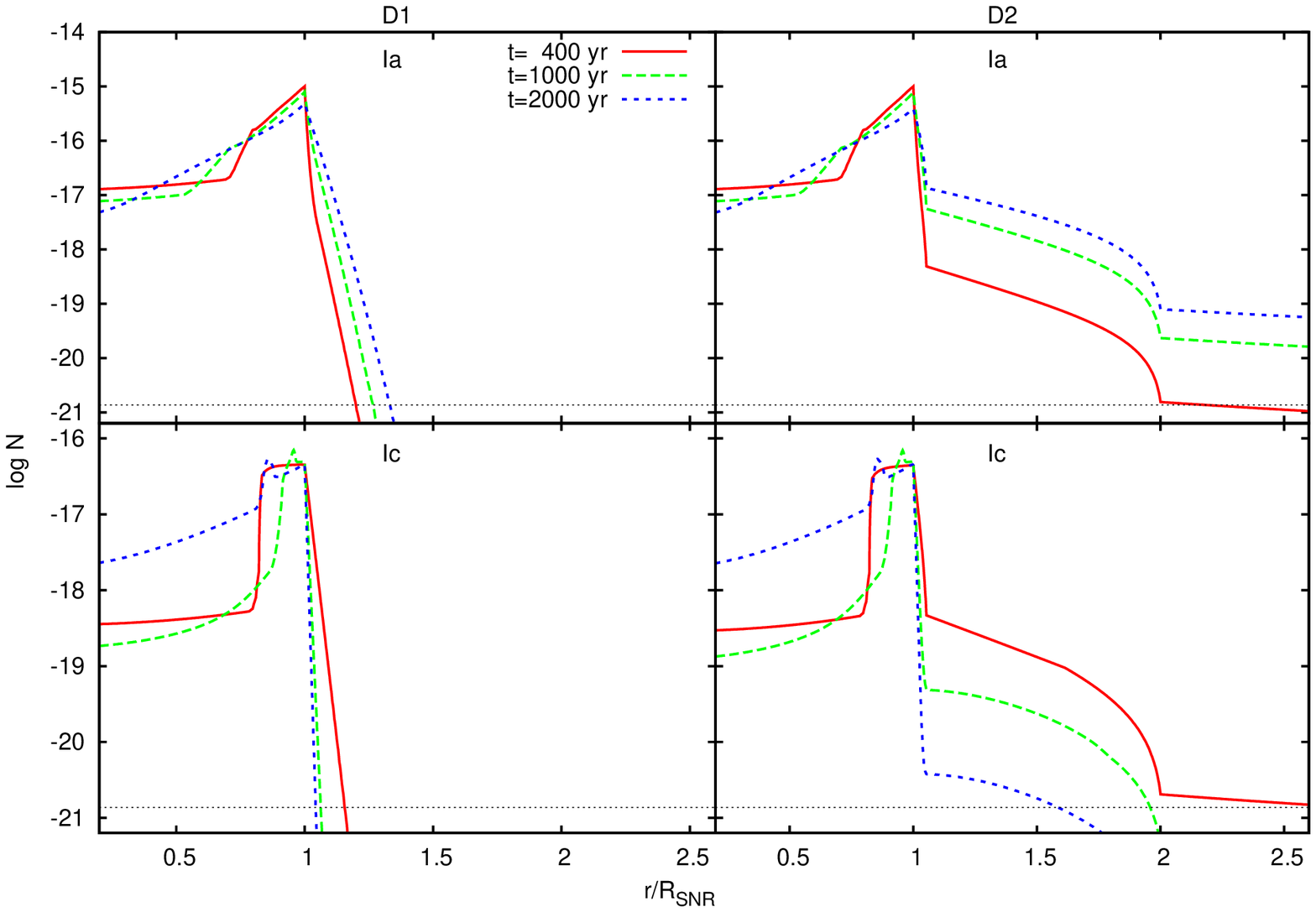}}
\caption{Radial distribution of CRs with energy 20~TeV at
different times compared with the Galactic CR background 
at the same energy (dotted line) for diffusion models
``D1'' (left) and ``D2'' (right). Type-Ia SNRs are shown in the top
row and core-collapse in the bottom row.}
\label{nar20}
\end{figure*}

\subsection{Spatial distributions of CRs}

\begin{figure*}[!t]
\centerline{\includegraphics[width=0.95\textwidth]{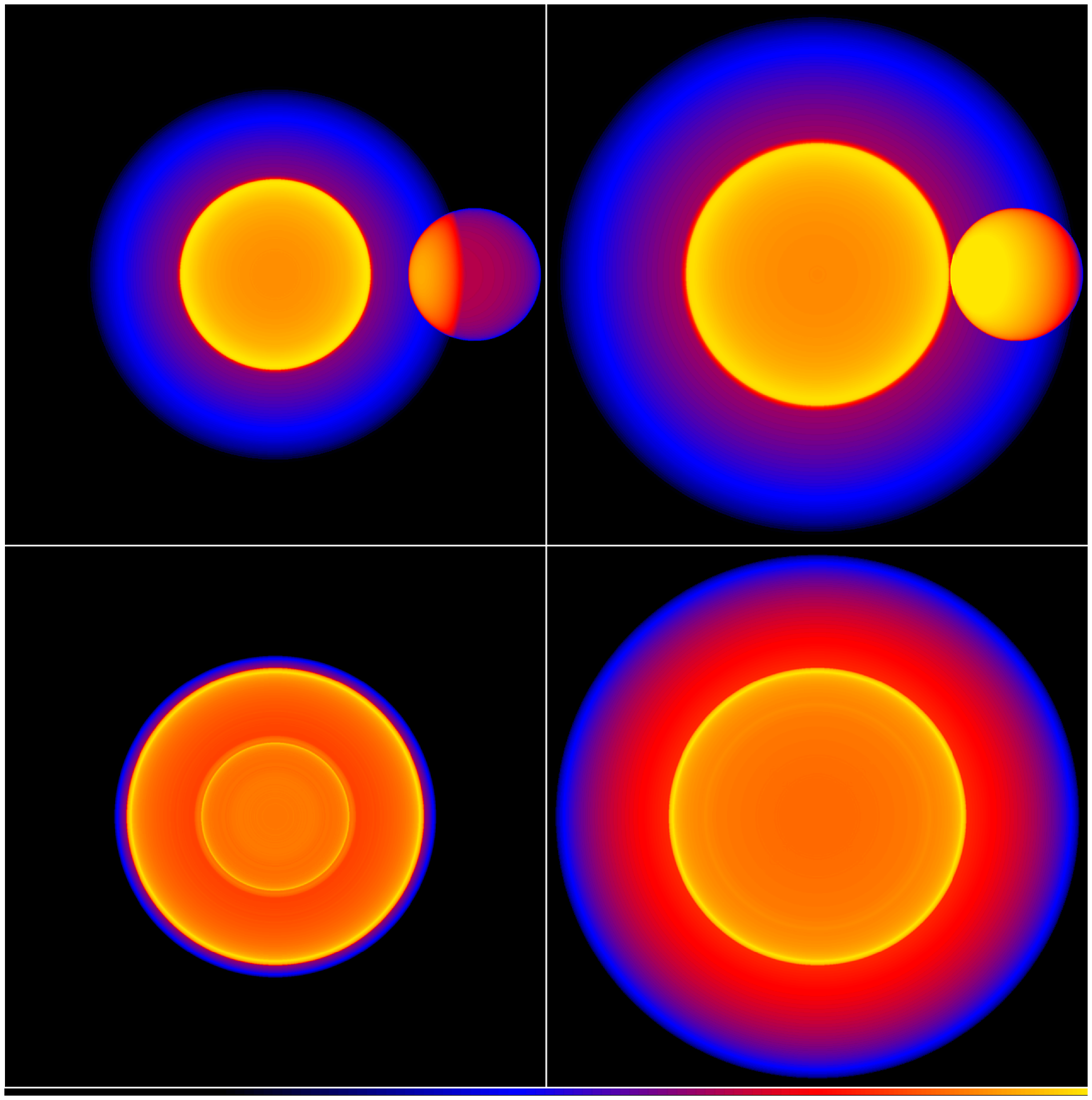}}
\caption{1-TeV Intensity distributions of type-Ia/MC (top) and
  CC-SNR/shell (bottom) at the age of 1000 (left) and 2000 (right)
  years. The log-scaled colormap spans roughly over 2.5 orders of
  magnitude for each image.}
\label{br}
\end{figure*}

To study the propagation of escaping particles in the medium around
SNRs and understand the significance of their contribution to the
local CR level, we plot the radial distributions of particles of given
energies for different types of SNR and different diffusion
model. Fig.~\ref{nar} presents the radial distributions of protons at
the peak energy in the spectra of escaped CRs, which is sometimes referred to as
$E_{\rm max}$. This energy changes with time, therefore at each time
particles of different energy are plotted (thick lines), along with
the corresponding Galactic background (thin lines). The evolution of
$E_{\rm max}$ is shown in Fig.~\ref{pmax} and discussed below. We
observe significant confinement of cosmic-ray particles close to the
shock for all SNRs and diffusion models, even at the highest particle
energies. Remarkably, for Type-Ia SNR in both diffusion models the
highest-energy CRs still dominate over the background at a few
SNR radii, whereas for CC-SNR this happens only in the D2 diffusion
model. This may be because in our simulations the
Type-Ia SNR propagates into a medium of higher average density, and
therefore has a radius roughly twice as small as the CC-SNR, which
implies less dilution. The CC-SNR produce a lower CR energy density
than Type-Ia SNRs owing to the lower density at the FS, hence lower
injection into the acceleration process. Therefore, the intensity of
CRs around CC-SNR only marginally exceeds the background. While these
results are applicable to Type-Ic SNR, they are not to SNR of Type-IIP.
The latter could have even smaller radii than Type-Ia, for
instance, as they expand within the high-density red supergiant wind. In
future papers, we will discuss the various SNR types in more detail.

We are particularly interested in the spatial distribution of CRs that produce
TeV-band gamma-ray emission.  We show the time evolution of these
distributions for a CR energy of 20~TeV in Fig.~\ref{nar20}, including
for comparison the Galactic CR background. We note from this figure
that in the D1 diffusion model, 20-TeV CRs are trapped very close to
the SNR shock ($\lesssim 1.4 R_{SNR}$ for Type-Ia SNR and $\lesssim
1.2 R_{SNR}$ for CC-SNR) at all times. Even in the D2 model, in which the diffusion
coefficient is very large, at late times the
intensity of 20-TeV CRs exceeds the background at a few $R_{SNR}$ only
in the case of a Type-Ia SNR.  Therefore, the detection of gamma-ray emission from MCs as probe of hadronic acceleration in SNR in the era of \textit{Fermi} and
current CTs is a far more complicated task than
previously thought. The object observed is always a composite of the SNR itself,
emission from the MC, or a dense shell illuminated by freshly
accelerated CRs, and the emission of interactions between upstream CRs and the
ambient diluted gas. One may be successful in scenarios in which the
MC is located close to the SNR. The distance to the MC
should then be generally smaller or comparable to $R_{SNR}$, otherwise at the location of the MC/shell
the background CRs dominate over the CRs from the SNR. This is a direct consequence of
both diffusion and dilution, which becomes a serious factor at large
distances from the SNR.

\subsection{Gamma-ray emission from the SNR-MC/shell}

We calculated the gamma-ray emission from the MC/shell
illuminated by CRs that had escaped from the SNR. For comparison, we plot their
spectra together with the emission spectra from the SNRs themselves
(Fig.~\ref{T1} and Fig.~\ref{T2}, lower panels). The gamma-ray
emission is only moderately affected by the choice of diffusion model. In the
case of the D2 model, we observe higher fluxes at lower energies
and also that the spectral peak is shifted to lower energies, thus making the
spectra soft in the band above $\sim 10$~TeV. The spectra thus reflect
the spatial variation in the diffusion coefficient in the vicinity of
the SNR. In Type-Ia SNR, the MC is always significantly brighter than
expected when illuminated by Galactic CRs. In contrast, the dense
shell around a CC-SNR does not experience a significantly enhanced CR
illumination until the shock is very close. The emission from the
CC-SNR itself is never particularly bright, and so at late times the
shell emission strongly dominates. On the other hand, the Type-Ia SNR 
dominates the intensity distribution of the SNR-MC system in all
scenarios and diffusion models, but only at the latest time, and in the
``near'' scenario the emission levels of SNR and MC become
comparable. The intensity of gamma-ray emission from the
MC/shell obviously strongly depends on the mass carried by the MC/shell,
but the trend described above holds and may permit us to differentiate
between Type-Ia and Type-Ic SNRs .

Additionally, for model D2 we constructed the intensity
distributions of gamma-ray emission from Type-Ia SNR/MC and
CC-SNR/shell systems, along with the background created by the
upstream CRs in the vicinity of the SNR (see Fig.~\ref{br}).
Since the number density of CRs confined within boundary $L$
significantly exceeds the number density of CRs beyond $L$, we
consider $L$ as the boundary of the SNR vicinity.  One can see how the
expanding halo of upstream CRs illuminates the surrounding
diluted gas and dense matter. The sharp contrast in brightness between
the inner and outer parts of the MC at the age of 1000 years arises
from the abrupt transition from intermediate to Galactic diffusion and
is in that sense an artifact. Moreover, as noted earlier in
  subsection~\ref{pspe}, if the diffusion coefficient inside the MC were
  larger than that in the surrounding medium, the CR illumination of
  the MC after 1000~years would be more uniform so that the far side
  becomes brighter. However, the image, though idealized, is not
unrealistic. The bright outer ring in CC-SNR intensity maps is the
swept-up shell illuminated by escaping CRs.  The smaller ring at the
age of 1000 years corresponds to the contact discontinuity (CD) of the
CC-SNR. The FS can almost not be seen owing to the very low density in the
wind zone. At the age of 2000 years, the shell strongly dominates, and
the shocks are not seen. There is only a marginal hint of emission
from the CD region.

\subsection{$E_{\rm max}$}

Different authors may define $E_{\rm max}$ in different ways, e.g.,
the maximum energy at which particles leave the system, the cut-off
energy, or the peak in the spectrum of escaped particles. In practice,
these definitions should provide nearly identical values. In some
models of CR acceleration, including analytical studies of CR escape,
$E_{\rm max}$ is a free parameter used in lieu of a free-escape
boundary.

The time evolution of $E_{\rm max}$ is a crucial issue. We derive
$E_{\rm max}$ from fitting the spectra of escaped CRs for each SNR
type and diffusion models.  The results are plotted in Fig.~\ref{pmax},
along with the results of \citet{EllByk11} and \citet{Gabetal09}. Although
we consider MF amplification (MFA), we do not find as
rapid a decrease in $E_{\rm max}$ as predicted by \citet{Gabetal09}. Our curves are more compatible with those of
\citet{EllByk11}, although they did not
include MFA. Most analytical studies are based on the Sedov solution,
MFA, and the assumption that particles start leaving the system at the
beginning of the Sedov stage when the highest energies are reached
\citep[e.g.][]{Gabetal09}. When we accurately account for the
evolution of a young SNR, in which the shock velocity is not constant
and falls with time, and we include the effects of dilution by considering
spherically symmetric systems, we find that particles leave the SNR
during the ``free expansion'' stage. In addition, \citet{Dru11}
reminds us that escape is a random process and it is only in a probabilistic sense controlled by the ratio of the diffusion length to SNR radius. Therefore, in our opinion solving one diffusion-advection
equation for all CRs, both those trapped inside the SNR and
those ``escaping'', is the best way to treat the problem. At
any instant, one can easily see which particles are inside and which
outside the SNR. Defining $E_{\rm max}$ by the spectrum of
CRs that have escaped into the MC/shell and observing the time
evolution of $E_{\rm max}$, one notices a dramatic difference between our result and that derived with
Sedov scaling. Moreover, as shown in Fig.~\ref{pmax}, besides the
early SNR evolution and SNR type, the evolution of $E_{\rm max}$ may
also depend on the assumed diffusion model. We must note, however,
that the trend seen in CC-SNRs for a D2 diffusion model may be an
artifact arising from our choice of transition boundary from Bohm to
intermediate diffusion coefficient, $\chi D_G$. In our case, the
transition is at 5\% of the SNR radius, which given the CC-SNR
evolution, is roughly equal to the size of the CR precursor,
$D_B(E_{\rm max})/V_s$. If the CR precursor extends into a region with a
large diffusion coefficient, shock acceleration loses efficiency, thus
effectively prescribing the $E_{\rm max}$. With time and increasing SNR
radius, the physical separation of the forward shock and the
transition point in the diffusion model increases, thus permitting
particles to reach higher energies.

\subsection{General remarks}

\begin{figure}[!t]
\centerline{\includegraphics[width=0.49\textwidth]{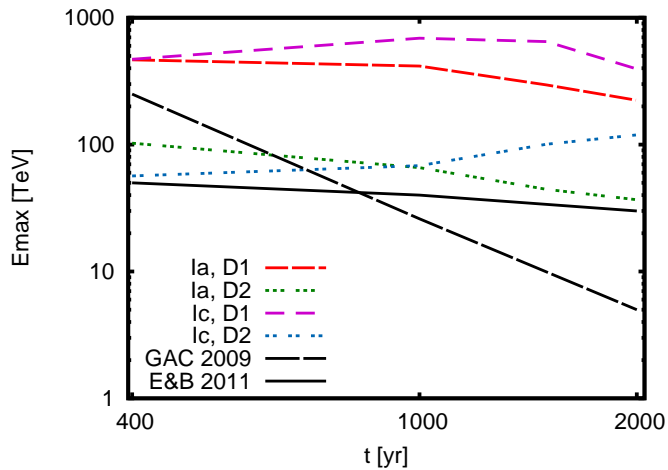}}
\caption{Time evolution of $E_{\rm max}$, the peak energy in the spectrum of
escaped CRs.}
\label{pmax}
\end{figure}

\citet{EllByk11} provided analytical approximations of the spectra of
escaping particles, which were then incorporated into their Monte Carlo
cosmic-ray propagation model. Our results on the spectra of escaped
CRs are compatible with those of \citet{EllByk11}, if the shock is sufficiently
far away from the location of interest and the system is still young.  At
later times, the time evolution of $E_{\rm max}$ (see Fig.~\ref{pmax})
leads to the formation of tails in the spectra of escaped CRs, that
are populated by particles that left the remnant early (see the 2000-year
plot at Fig.~\ref{T2}).

We noted that for both types of SNR and diffusion model
(especially in D1) we find a significantly smaller intensity ratio of
escaped CRs to trapped CRs at $E_{\rm max}$ than \citet{EllByk11},
even considering their use of NDSA. This discrepancy possibly arises
from the neglect of dilution in their model.

We compared the radial profiles with those presented in the
literature. In general, the radial profiles obtained in our D2 models
are compatible with those in \citet{Fujetal11} (model C disregarding
the growth of Alfv\'en waves) and with the profiles of \citet{EllByk11}
(propagation in pure CSM) out to a radius of $R_{\rm SNR}$ from the
shock. These authors, however, did not make any comparisons with the
background level as their results are presented in arbitrary units. We
were unable to find research that shows the profiles for pure
Bohm diffusion out to a full SNR radius from the shock.

\subsection{Observational implications} 

The TeV-band intensity map of the young Type-Ia SNR Tycho observed
with VERITAS \citep{Accetal11} shows off-center emission, which may
correspond to a fluctuation. However, it may also indicate an interaction of escaping
CRs with a nearby MC. If this is true, Fig.~\ref{T1} suggests
that in the GeV band emission from the SNR itself should dominate. So, if
the TeV-band emission indeed partly originates from an illuminated MC outside
the remnant, the component from the SNR itself must have a softer
spectrum than suggested by a naive comparison of \textit{Fermi} and
VERITAS data.

Another interesting example is the young TeV-bright CC-SNR
RX~J1713.7-3946 detected with \textit{Fermi}
\citep{Abdetal11_RX1713}.  \citet{Elletal11} applied their model to
explain its high-energy emission via escaping CRs and found that this
scenario is very unlikely. However, Fig.~\ref{T2}
indicates peculiar spectral transitions, in particular for the D2
model, that make multi-component spectra quite conceivable in which
most of the GeV-band emission arises from the SNR itself, whereas a
substantial fraction of TeV-band emission originates from gas exterior to the
remnant. More careful modeling is needed to test the applicability of
this scenario to RX~J1713.7-3946.

Although our treatment extends only over the first 2000 years of SNR
evolution, we note that the rather peculiar spectra of CC-SNRs
found in our models have spectral breaks that appear similar to those
observed from some older \textit{Fermi} SNRs \citep[e.g.][]{CasSla10,
  Abdetal10_ic443}. In addition, the observed discrepancy in the
  locations of the TeV and (sub-)GeV emission in the SNR vicinity can
  be considered in line with our models. For instance, as seen from the
  lower panels of Fig.~\ref{T1} and~\ref{T2}, the emission from the
  dense matter far away from the SNR may dominate at TeV energies and
  disappear at low (sub-)GeV energies. At the same time, the emission
  from the SNR itself would be stronger at GeV and weaker at TeV
  energies. Altogether, this makes the emission ``peak'' move over the
  image with the changing energy of the observations. In some sense, a similar
  scenario is put forward by the AGILE team \citep{Guietal11} for the
  regions around SNRs W28 and IC~443, to explain why in
  the GeV-band the peak-emission error box is separated from the
  TeV-emission error box observed by VERITAS and MAGIC.

\section{Conclusions}

We have modified our modeling technique developed in \cite{Teletal11}
to examine the distribution of escaped CRs at distances far from
SNRs. Our approach combines a realistic treatment of SNR evolution
using hydrodynamic simulations with test-particle calculations of CR
acceleration, by solving the transport equation of cosmic rays in a
spherically symmetric geometry. Both the reverse and forward
shocks are included in these simulations. We also account for the
re-acceleration at reflected shocks in core-collapse SNR. Our
approach inherently includes the effects of dilution due to the
expansion of the system, as well as the finite size of the SNR and any
gas target illuminated by escaping CRs. It permits the reconstruction of the
spectra of escaped CRs at any given distance from the source.

We have shown that the peak energy and intensity of escaped particles
strongly depends on the efficiency of the diffusion in the vicinity
of the SNR. If diffusion is Bohmian out to the boundary $L = 2 R_{\rm
  SNR}$, CRs are very efficiently confined to the SNR, and only the
highest-energy particles are able to diffuse out to some distance
  from the SNR. Even though CRs at $E_{\rm max}$ leave the
SNR, they are still trapped at distances far less than one SNR radius
from the FS. If the diffusion coefficient in the FS upstream region is
much larger than Bohmian, but not at the level of average Galactic
diffusion, then CRs of lower energy are able to escape from the SNR.
Consequently, one observes the broader spectra of escaped CRs with lower
$E_{\rm max}$ and the cut off in the spectra of confined CRs is slower
than exponential.

It is possible but difficult to constrain the
diffusion coefficient in the vicinity of the SNR using the emission
from nearby dense material. The particles cannot generally
propagate far from the shock. Thus, to study CR acceleration in SNRs
via CR escape into a nearby MC or shell, one needs to find MCs or shells
located very close, unless the diffusion coefficient in the immediate
environment of the SNR is a significant fraction of the average
Galactic one, in which case $E_{\rm max}$ is lower. We note that in
all diffusion models, and especially in the ``far``
scenarios, one can clearly observe the striking effect of dilution,
i.e. the decrease in the CR intensity upstream of the FS, introduced by
the spherical geometry of the SNR.  Ignoring the spherical geometry of
the system is a serious mistake when one models the gamma-ray emission
from MCs placed at significant distances from the SNRs.

We find that Type-Ia SNR in our simulation has a higher CR energy
density in its vicinity than Type-Ic core-collapse SNR. Therefore, Type-Ia
SNR may have brighter pion-decay gamma-ray emission than Type-Ic SNR
that is still evolving in the wind zones of its progenitor
star. The emission of Type-Ia SNR is normally brighter than
the MC emission until the SNR shock is very close to the MC. Only
then is the gamma-ray flux from the MC comparable to that from
the SNR. The emission from the CC-SNR and the wind-blown shell is initially near the
background level. At a later stage, the shell emission can significantly
dominate over the SNR emission. Although the exact scenario depends
on the MC/shell parameters, most importantly the target mass, the
described trend may serve to differentiate between these SNR types.

Finally, we find that the commonly used approximation for the $E_{\rm max}$
evolution based on Sedov scaling is not reproduced when the
evolution of the SNR shocks is computed accurately. Moreover, the
$E_{\rm max}$ behavior critically depends on the diffusion coefficient
in the vicinity of the SNR.

\acknowledgements 

VVD's research is supported by NASA through Chandra awards issued by
the Chandra X-ray Observatory Center and NASA/Fermi grant NNX10AO44G.

\bibliographystyle{aa}
\bibliography{escape}

\end{document}